\documentclass[spanish,12pt]{article}
\usepackage{amssymb}
\input{tcilatex}
\newif\ifpdf
\ifx\pdfoutput\undefined
 \pdffalse
 \usepackage{graphicx}
\else
 \usepackage[pdftex]{graphicx}
 \usepackage{times}
 \pdftrue
 \pdfoutput=1          
\fi

\newcommand{\pict}[6]{%
 \begin{figure}[!ht]\begin{center}
  \ifpdf
    \includegraphics[#3]{#4}
  \else
    \includegraphics[#1]{#2}
  \fi
  \caption{#5}
  \label{#6}
 \end{center}\end{figure}
}

\begin{document}

\title{Correspondence  Analysis for Symbolic Multi--Valued Variables}

\author{ Oldemar Rodr\'{\i}guez \thanks{University of Costa Rica, San Jos\'e, Costa Rica;
E-Mail: oldemar.rodriguez@ucr.ac.cr}}

\date{Recibido: 10 de marzo 2007}

\maketitle

\begin{abstract}
This paper sets a proposal of a new method and two new algorithms
for Correspondence Analysis when we have Symbolic Multi--Valued
Variables (SymCA). In our method, there are two multi--valued
variables $X$ and $Y$, that is to say, the modality that takes the
variables for a given individual is a finite set formed by the
possible modalities taken for the variables in a given individual,
that which allows to apply the Correspondence Analysis to multiple
selection questionnaires. Then, starting from all the possible
classic contingency tables an interval contingency table can be
built, which will be the point of departure of the proposed method.
\end{abstract}

Symbolic data analysis, contingency tables,  interval contingency table,
disjunctive complete table,
symbolic multi--valued variables.

\section{Relationship between two symbolic multi--valued variables}



One of the objectives of the Correspondence Analysis is to explore
the relationships between the modalities of two qualitative
variables by two--dimensional representations. The objective of the
Correspondence Factorial Analysis between two symbolic multi--valued
variables (SymCA) is the same one, but also there is some kind of
uncertainty in the input data that will be reflected in the
two--dimensional representations, since each modality will be
represented with a rectangle instead of a point, as usual. A
symbolic variable $Y$ is called multi--valued one if its values
$Y(k)$ are all finite subsets [Bock and Diday (2000)].

In the classic Correspondence Analysis a contingency table
associated with two qualitative variables is build. For example,
suppose there are two qualitative variables: $X=$eyes-color (with 3
modalities green, blue and brown) and $Y=$ hair-color (with 2
modalities blond and black). If each one of the 2 qualitative
variables is observed in 5 individuals the following disjunctive
complete tables could be obtained:

\[
X=\left[
\begin{array}{ccc}
1 & 0 & 0 \\
0 & 0 & 1 \\
1 & 0 & 0 \\
0 & 0 & 1 \\
1 & 0 & 0
\end{array}
\right] ,\qquad Y=\left[
\begin{array}{cc}
0 & 1 \\
0 & 1 \\
0 & 1 \\
1 & 0 \\
1 & 0
\end{array}
\right] .
\]

In these matrices, if the $(i,j)-$entry is 1, it means that the
individual $i$ take the modality $j$ and a 0 means that individual
$i$ doesn't take it. If the matrix multiplication $K=X^{t}Y$ is
made, the crossed table or contingency table between the variables
$X$ and $Y$ will be obtained, such as the following:

\[
K=X^{t}Y=\left[
\begin{array}{cc}
1 & 2 \\
0 & 0 \\
1 & 1
\end{array}
\right] .
\]

In the $(i,j)-$entry of the matrix $K$ appears the quantity of
individuals that assume simultaneously the modality $i$ of the
variable $X$ and the modality $j$ of the variable $Y$. As it is well
known, the Correspondence Analysis usually starts up with the
contingency table among the variables $X$ and $Y$.

For the case of multi--valued variables there will be individuals
whose information regards the assumed modality is a ``diffuse
variable".

\textbf{Example 1.} Let $X$ be the qualitative variable
``\textit{eyes-color}'' with 3 modalities: green, blue and brown; it
could be that the eyes color of the first individual is green or
blue (but not both), that is to say $X(1)=$green or $X(1)=$blue. Let
$Y$ be the qualitative variable ``\textit{hair-color}'' with 2
modalities blond and black, there could also be an individual whose
hair color is not completely well defined, for instance for the
third individual, it might also be $Y(3)=$blond or $Y(3)=$black but
not both). In this way there are two possible disjunctive complete
tables for the variable $X$ and two for $Y$, which are:

\[
X_{1}=\left[
\begin{array}{ccc}
1 & 0 & 0 \\
0 & 0 & 1 \\
1 & 0 & 0 \\
0 & 0 & 1 \\
1 & 0 & 0
\end{array}
\right] ,\quad X_{2}=\left[
\begin{array}{ccc}
0 & 1 & 0 \\
0 & 0 & 1 \\
1 & 0 & 0 \\
0 & 0 & 1 \\
1 & 0 & 0
\end{array}
\right] ,\qquad Y_{1}=\left[
\begin{array}{cc}
0 & 1 \\
0 & 1 \\
1 & 0 \\
1 & 0 \\
1 & 0
\end{array}
\right] ,\quad Y_{2}=\left[
\begin{array}{cc}
0 & 1 \\
0 & 1 \\
0 & 1 \\
1 & 0 \\
1 & 0
\end{array}
\right] .
\]

Using this information there will be 4 possible contingency tables
among the variable $X$ and $Y$, which are:

\[
K_{1}=X_{1}^{t}Y_{1}=\left[
\begin{array}{cc}
2 & 1 \\
0 & 0 \\
1 & 1
\end{array}
\right] ,\quad K_{2}=X_{2}^{t}Y_{1}=\left[
\begin{array}{cc}
2 & 0 \\
0 & 1 \\
1 & 1
\end{array}
\right] ,
\]
\[
K_{3}=X_{1}^{t}Y_{2}=\left[
\begin{array}{cc}
1 & 2 \\
0 & 0 \\
1 & 1
\end{array}
\right] ,\quad K_{4}=X_{2}^{t}Y_{2}=\left[
\begin{array}{cc}
1 & 1 \\
0 & 1 \\
1 & 1
\end{array}
\right] .
\]

Taking the minimum and the maximum of the components of these 4
matrices, a contingency data table of interval type is obtained:

\[
K=\left[
\begin{array}{cc}
\lbrack 1,2] & [0,2] \\
\lbrack 0,0] & [0,1] \\
\lbrack 1,1] & [1,1]
\end{array}
\right] .
\]

The main idea of the  proposed method is to carry out a classic
Correspondence Analysis on the matrix of $K$'centers, working with a
similar idea just like in the Centers Method in principal component
analysis for interval data (see [Cazes and other 1997]).

The construction of the matrix $K$ of interval type requires many
calculations. If the variable $X$ has $p$ modalities taken in $m$
individuals then the row $i$ of the disjuntive complete table of $X$
has at most $p$ possibilities to place the value 1 (it cannot be at
the same time in any pair of components), then there are at most
$p^{m}$ possible disjunctive complete tables for the variable $X$.
Similarly, if the variable $Y$ has $n$ modalities and they are
observed in $m$ individuals then there are $n^{m}$ possible
disjunctive complete tables for the variable $Y$. Therefore there
are $p^{m}n^{m}$ possible contingency matrices associated to the
variables $X$ and $Y$. Then $K$ should be generated taking the
minimum and the maximum of these $p^{m}n^{m}$ matrices. That is to
say, $p^{m}n^{m}$ products of matrices of sizes $p\times m$ and
$m\times n$ should be made.

The following theorem reduces the matrix $K$ calculation to only two
matrix multiplications, therefore it has time $\mathcal{O}(m^{2})$
(extremely quick). Before presenting the theorem, the following
definition must be given.

\textbf{Definition 1.} Let $X$ be a qualitative multi--valued
variable. The matrix of minimum possibilities (meet matrix)
associated to $X$ is defined and denoted by $\underline{X}$, such as
the following:

\[
\underline{X}_{ij}=\left\{
\begin{array}{cl}
0 & \text{ if there is }j\neq j^{\prime }\text{ such that the
individual }i\text{ could take the modality }j \\
& \text{ or the }j^{\prime }\text{ of }X.\text{ }
\\
1 & \text{ if the individual }i\text{ only took the modality }j%
\text{ of }X\text{.}
\end{array}
\right.
\]

It could be said that:

\[
\underline{X}_{ij}=\left\{
\begin{array}{cl}
0 & \text{ if } X(i) \neq \{j\}\\
1 & \text{ if }X(i)=\{j\}.
\end{array}
\right.
\]

Also, the matrix of maximum possibilities (join matrix) associated
to $X$ is defined and denoted by $\overline{X}$, as follows:

\[
\overline{X}_{ij}=\left\{
\begin{array}{cl}
0 & \text{ if the individual }i\text{ doesn't take the modality }j\text{ of }X,%
\text{ } \\
1& \text{ if the individual }i\text{ can take the modality }j\text{ of }X%
\text{,}
\end{array}
\right.
\]

it could be said that:

\[
\overline{X}_{ij}=\left\{
\begin{array}{cl}
0 & \text{ if } j \notin X(i) \\
1 & \text{ if } j \in X(i).
\end{array}
\right.
\]

\textbf{Example 2.} Using the same variables $X$ and $Y$ from
example 1, it is obtained:

\[
\underline{X}=\left[
\begin{array}{ccc}
0 & 0 & 0 \\
0 & 0 & 1 \\
1 & 0 & 0 \\
0 & 0 & 1 \\
1 & 0 & 0
\end{array}
\right] ,\quad \overline{X}=\left[
\begin{array}{ccc}
1 & 1 & 0 \\
0 & 0 & 1 \\
1 & 0 & 0 \\
0 & 0 & 1 \\
1 & 0 & 0
\end{array}
\right] ,
\]
\[
\underline{Y}=\left[
\begin{array}{cc}
0 & 1 \\
0 & 1 \\
0 & 0 \\
1 & 0 \\
1 & 0
\end{array}
\right] ,\quad \overline{Y}=\left[
\begin{array}{cc}
0 & 1 \\
0 & 1 \\
1 & 1 \\
1 & 0 \\
1 & 0
\end{array}
\right] .
\]

\textbf{Theorem 1.} Let $X$ and $Y$ be two qualitative multi--valued
variables  and let $K_{ij}=\left [ \underline{k_{ij}},
\overline{k_{ij}} \right]$ the contingency matrix of interval type
associated to $X$ and $Y$, then:

\begin{eqnarray*}
\underline{k_{ij}} &=&\left( \underline{X}^{t}\underline{Y}\right) _{ij}%
\text{ for }i=1,2,\ldots ,p\text{ and }j=1,2,\ldots ,n. \\
\overline{k_{ij}} &=&\left( \overline{X}^{t}\overline{Y}\right)
_{ij}\text{ for }i=1,2,\ldots ,p\text{ and }j=1,2,\ldots ,n.
\end{eqnarray*}

\textbf{Proof:} It is evident that $\left (\underline{X}^{t}\underline{Y}%
\right) _{ij}$ counts the worst of the cases for the modality $i$ of
the variable $X$ and the modality $j$ of the variable $Y$, that is
to say, the minimum of individuals that take at the same time the
modality $i$ of the variable $X$ and the
modality $j$ of the variable $Y$. While $\left (\overline{X}^{t}%
\overline{Y}\right) _{ij}$ counts the best of the cases (the maximum
of individuals) for the modality $i$ of the variable $X$ and the
modality $j$ of the variable $Y$, that is to say, the maximum of
individuals that take at the same time the modality $i$ of the
variable $X$ and the modality $j$ of the variable $Y$.\hfill
$\blacksquare $

\textbf{Example 3.} Using the same variables $X$ and $Y$ as in
example 2, it is obtained:

\[
\underline{X}^{t}\underline{Y}=\left[
\begin{array}{ccccc}
0 & 0 & 1 & 0 & 1 \\
0 & 0 & 0 & 0 & 0 \\
0 & 1 & 0 & 1 & 0
\end{array}
\right] \left[
\begin{array}{cc}
0 & 1 \\
0 & 1 \\
0 & 0 \\
1 & 0 \\
1 & 0
\end{array}
\right] =\left[
\begin{array}{cc}
1 & 0 \\
0 & 0 \\
1 & 1
\end{array}
\right] ,
\]
\[
\overline{X}^{t}\overline{Y}=\left[
\begin{array}{ccccc}
1 & 0 & 1 & 0 & 1 \\
1 & 0 & 0 & 0 & 0 \\
0 & 1 & 0 & 1 & 0
\end{array}
\right] \left[
\begin{array}{cc}
0 & 1 \\
0 & 1 \\
1 & 1 \\
1 & 0 \\
1 & 0
\end{array}
\right] =\left[
\begin{array}{cc}
2 & 2 \\
0 & 1 \\
1 & 1
\end{array}
\right] ,
\]

Then:

\[
K=\left[
\begin{array}{cc}
\lbrack 1,2] & [0,2] \\
\lbrack 0,0] & [0,1] \\
\lbrack 1,1] & [1,1]
\end{array}
\right] .
\]

\section{Correspondence factorial analysis between two symbolic
multi--valued variables}

In the method proposed, there are two multi--valued variables $X$
and $Y$, that is to say, the modality that takes the variables for a
given individual is a finite set formed by the possible modalities
taken for the variables in a given individual. As we explained in
the previous section, starting from the classic contingency tables
an interval contingency table can be built, which will be the point
of departure of the method proposed.

An interval contingency matrix $K$ with $n$ rows and $p$ columns
associated to two multi--valued variables $X$ and $Y$ is taken into
consideration, where $X$ has $n$ modalities and $Y$ has $p$
modalities.

\begin{equation}
K=\left(
\begin{array}{ccc}
\left[ \underline{k_{11}},\overline{k_{11}}\right] & \cdots & \left[
\underline{k_{1p}},\overline{k_{1p}}\right] \\
\vdots & \ddots & \vdots \\
\left[ \underline{k_{n1}},\overline{k_{n1}}\right] & \cdots & \left[
\underline{k_{np}},\overline{k_{np}}\right]
\end{array}
\right) .  \label{m1}
\end{equation}

The idea of the method is to transform the matrix presented in
(\ref{m1}) into the following matrix (\ref{m2}):

\begin{equation}
K^{c}=\left(
\begin{array}{cccc}
k_{11}^{c} & k_{12}^{c} & \cdots & k_{1p}^{c} \\
k_{21}^{c} & k_{22}^{c} & \cdots & k_{2p}^{c} \\
\vdots & \vdots & \ddots & \vdots \\
k_{n1}^{c} & k_{p2}^{c} & \cdots & k_{np}^{c}
\end{array}
\right) =\left(
\begin{array}{cccc}
\frac{\underline{k_{11}}+\overline{k_{11}}}{2} & \frac{\underline{k_{12}}+%
\overline{k_{12}}}{2} & \cdots & \frac{\underline{k_{1p}}+\overline{k_{1p}}}{%
2} \\
\frac{\underline{k_{21}}+\overline{k_{21}}}{2} & \frac{\underline{k_{22}}+%
\overline{k_{22}}}{2} & \cdots & \frac{\underline{k_{2p}}+\overline{k_{2p}}}{%
2} \\
\vdots & \vdots & \ddots & \vdots \\
\frac{\underline{k_{n1}}+\overline{k_{n1}}}{2} & \frac{\underline{k_{n2}}+%
\overline{k_{n2}}}{2} & \cdots & \frac{\underline{k_{np}}+\overline{k_{np}}}{%
2}
\end{array}
\right) .  \label{m2}
\end{equation}

A classic Correspondence Analysis of the matrix $K^{c}$ is carried
out to do that, we make a PCA (Principal Component Analysis) of row
profiles and column profiles of $K^{c}$, that allows to obtain
two-dimensional representations of the centers. Then, in a similar
way to the method of the centers in PCA for interval data, the tops
of the hypercubes are projected in supplementary in this plane, then
we choose the minimum and the maximum. The difference in this case
is that row hypercubes and column hypercubes are projected in the
same plane (simultaneous representation).

For this, as it is usual in Correspondence Analysis, the following
notation is introduced:

\[
k_{i\cdot }^{c}=\dsum\limits_{j=1}^{p}k_{ij}^{c},\quad k_{\cdot
j}^{c}=\dsum\limits_{i=1}^{n}k_{ij}^{c},\quad
k^{c}=\dsum\limits_{i=1,j=1}^{n,p}k_{ij}^{c},
\]

and the ``relative frequencies'':

\[
f_{ij}^{c}=\frac{k_{ij}^{c}}{k^{c}},\quad f_{i\cdot
}^{c}=\dsum\limits_{j=1}^{p}f_{ij}^{c},\quad f_{\cdot
j}^{c}=\dsum\limits_{i=1}^{n}f_{ij}^{c}.
\]

In the classic Correspondence Analysis, to analyze a contingency
table  the effective table is not used, but the table of row
profiles and column profiles are (that is to say, there is an
interest in the percentage distributions to the interior of the rows
and columns). The $i-$th row profile is defined by:

\[
\left(\frac{f_{ij}^{c}}{f_{i\cdot
}^{c}}\right)_j=\left(\frac{k_{ij}^{c}}{k_{i\cdot }^{c}}\right)_j,
\]

and the $j-$th column profile by:

\[
\left(\frac{f_{ij}^{c}}{f_{\cdot
j}^{c}}\right)_i=\left(\frac{k_{ij}^{c}}{k_{\cdot j}^{c}}\right)_i.
\]

In this way in the classic Correspondence Analysis in a simultaneous
way are represented the $n$ row profiles in $\Bbb{R}^{p}$ given by:

\[
\left\{ \frac{f_{ij}^{c}}{f_{i\cdot }^{c}}\text{ for }j=1,2,\ldots
,p\right\} ,
\]

and the $p$ column profiles in $\Bbb{R}^{n}$ given by:

\[
\left\{ \frac{f_{ij}^{c}}{f_{\cdot j}^{c}}\text{ for }i=1,2,\ldots
,n\right\} .
\]

Then the data table suffers two transformations, one on the row
profiles and another one on the column profiles, using that data
table clouds of points in $\Bbb{R}^{p}$ and in $\Bbb{R}^{n}$ will be
built. These transformations can be described in terms of three
matrices $F^{c},D_{n}^{c}$ and $D_{p}^{c}$. Where
$F^{c}=\frac{1}{k^{c} }K^{c}$ is a matrix of size $n\times p$ of the
relative frequencies ($\left ( F^{c}\right) _{ij}=f_{ij}^{c}$),
$D_{n}^{c}$ is a diagonal matrix of size $n\times n$ whose diagonal
is formed by those marginal of the rows $f_{i\cdot}^{c}$ and the
matrix $D_{p}^{c}$ is a diagonal matrix of size $p\times p$ whose
diagonal is formed by the marginal of the columns $f_{\cdot j}^{c}$.

With this notation, in the space $\Bbb{R}^{p}$ the row profiles that
are the rows of the matrix $X=\left(D_{n}^{c}\right)^{-1}F^{c}$ with
the metric $M=\left(D_{p}^{c}\right)^{-1}$ are represented, in which
case the distance between two row profiles is:

\[
d^{2}(i,k)=\sum_{j=1}^{p}\frac{1}{f_{\cdot j}^{c}}\left( \frac{f_{ij}^{c}}{%
f_{i\cdot }^{c}}-\frac{f_{kj}^{c}}{f_{k\cdot }^{c}}\right) ^{2},
\]

and in the space $\Bbb{R}^{n}$ the column profiles are represented
and constitute the columns of the matrix $X=\left
(D_{p}^{c}\right)^{-1}\left (F^{c}\right)^{t}$ with the metric
$M=\left(D_{n}^{c}\right)^{-1}$, in which case the distance between
two column profiles is:

\[
d^{2}(j,s)=\sum_{i=1}^{n}\frac{1}{f_{i\cdot }^{c}}\left( \frac{f_{ij}^{c}}{%
f_{\cdot j}^{c}}-\frac{f_{js}^{c}}{f_{\cdot s}^{c}}\right) ^{2}.
\]

As it is well known, to make this representation in the space
$\Bbb{R}^{p}$ the singular value decomposition of the matrix
$S^{c}=\left (F^{c}\right)^{t}\left (D_{n}^{c}\right)^{-1}F^{c}\left
(D_{p}^{c}\right)^{-1}$ must be done in such way that the factorial
coordinates are:

\[
\psi _{\alpha }=\left( D_{n}^{c}\right) ^{-1}F^{c}\left(
D_{p}^{c}\right) ^{-1}u_{\alpha },
\]

where $u_{\alpha}$ is the eigenvector of $S^{c}$ associated to the
eigenvalue $\lambda _{\alpha}$. Explicitly the factorial coordinates
in the space $\Bbb{R}^{p}$ are:

\begin{equation}
\psi _{\alpha i}=\sum_{j=1}^{p}\frac{f_{ij}^{c}}{f_{i\cdot
}^{c}f_{\cdot j}^{c}}u_{\alpha j}.  \label{eq11}
\end{equation}

It is also very well--known that to make this representation in the
space $\Bbb{R}^{n}$ we do the singular value decomposition of the
matrix $T^{c}=F^{c}\left(D_{p}^{c}\right)^{-1}\left
(F^{c}\right)^{t}\left(D_{n}^{c}\right)^{-1}$ in a such way that the
factorial coordinates are:

\[
\varphi _{\alpha }=\left( D_{p}^{c}\right) ^{-1}\left( F^{c}\right)
^{t}\left( D_{n}^{c}\right) ^{-1}v_{\alpha },
\]

where $v_{\alpha}$ is the eigenvector of $T^{c}$ associated to the
eigenvalue $\lambda _{\alpha}$ (the first eigenvalue is 1, so it is
discarded). Explicitly, the factorial coordinates in the space
$\Bbb{R}^{n}$ are:

\begin{equation}
\varphi _{\alpha j}=\sum_{i=1}^{n}\frac{f_{ij}^{c}}{f_{i\cdot
}^{c}f_{\cdot j}^{c}}v_{\alpha i}.  \label{eq12}
\end{equation}

The following two theorems will allow to project in form of
rectangles the column profiles and row profiles of interval type.
Where, if we denote $\underline{f_{ij}}=\dfrac{\underline{k_{ij}}}{k^{c}%
}$ and $\overline{f_{ij}}=\dfrac{\overline{k_{ij}}}{k^{c}}$ the
column profile of interval type is:

\[
\left[ \frac{\underline{f_{ij}}}{f_{\cdot j}^{c}},\frac{\overline{f_{ij}}}{%
f_{\cdot j}^{c}}\right] \text{ for }i=1,2,\ldots ,n,
\]

and the row profile of interval type  is:

\[
\left[ \frac{\underline{f_{ij}}}{f_{i\cdot }^{c}},\frac{\overline{f_{ij}}}{%
f_{i\cdot }^{c}}\right] \text{ for }j=1,2,\ldots ,p.
\]

\textbf{Theorem 2.} If the hypercube defined by the $j$--th column
profile of interval type is projected on the $\alpha$--th principal
component of the Correspondence Analysis of the matrix $K^{c}$ (in
the direction of $v_{\alpha}$), then the maximum and minimum values
are given by the equations (\ref{eq3}) and (\ref{eq4}) respectively.

\begin{equation}
\underline{\varphi _{\alpha j}}=\dsum\limits_{i=1,v_{\alpha i}<0}^{n}\frac{%
\overline{f_{ij}}}{f_{i\cdot }^{c}f_{\cdot j}^{c}}v_{\alpha
i}+\dsum\limits_{i=1,v_{\alpha
i}>0}^{n}\frac{\underline{f_{ij}}}{f_{i\cdot }^{c}f_{\cdot
j}^{c}}v_{\alpha i},  \label{eq3}
\end{equation}

\begin{equation}
\overline{\varphi _{\alpha j}}=\dsum\limits_{i=1,v_{\alpha i}<0}^{n}\frac{%
\underline{f_{ij}}}{f_{i\cdot }^{c}f_{\cdot j}^{c}}v_{\alpha
i}+\dsum\limits_{i=1,v_{\alpha
i}>0}^{n}\frac{\overline{f_{ij}}}{f_{i\cdot }^{c}f_{\cdot
j}^{c}}v_{\alpha i}.  \label{eq4}
\end{equation}

\textbf{Proof:} Let $z_{j}=(z_{1j},z_{2j},\ldots ,z_{nj})\in
Z_{H}^{j}$ be the hypercube defined by the $j$--th column profile of
interval type
$Y_{ij}=\left[ \dfrac{\underline{f_{ij}}}{f_{\cdot j}^{c}},\dfrac{\overline{f_{ij}}}{%
f_{\cdot j}^{c}}\right]$, then if $z_{ij}\in Y_{ij}$ for all
$i=1,2,\ldots ,n$ and $j=1,2,\ldots ,p$, we have that
$\dfrac{\underline{f_{ij}}}{f_{\cdot j}^{c}} \leq z_{ij} \leq
\dfrac{\overline{f_{ij}}}{f_{\cdot
 j}^{c}}$, then $z_{ij}=\dfrac{\widehat{z}_{ij}}{f_{\cdot j}^{c}}$
 with $\underline{f_{ij}} \leq \widehat{z}_{ij} \leq
 \overline{f_{ij}}$.
As $\dfrac{1}{f_{i\cdot }^{c}}>0$ we have:

\begin{equation}
\frac{\underline{f_{ij}}}{f_{i\cdot }^{c}f_{\cdot j}^{c}}v_{\alpha
i}\leq \frac{\widehat{z}_{ij}}{f_{i\cdot }^{c}f_{\cdot
j}^{c}}v_{\alpha i}\leq \frac{\overline{f_{ij}}}{f_{i\cdot
}^{c}f_{\cdot j}^{c}}v_{\alpha i}\text{ if }v_{\alpha i}\geq 0,
\label{eq:eqd71}
\end{equation}

and

\begin{equation}
\frac{\underline{f_{ij}}}{f_{i\cdot }^{c}f_{\cdot j}^{c}}v_{\alpha
i}\geq \frac{\widehat{z}_{ij}}{f_{i\cdot }^{c}f_{\cdot
j}^{c}}v_{\alpha i}\geq \frac{\overline{f_{ij}}}{f_{i\cdot
}^{c}f_{\cdot j}^{c}}v_{\alpha i}\text{ if }v_{\alpha i}\leq 0.
\label{eq:eqd72}
\end{equation}

Let by $pz_{\alpha j}$ the projection in supplementary of $z_{j}$ on
the factorial axis with direction $v_{\alpha}$. As it is very well
known in classic Correspondence Analysis, the projection in
supplementary has the form $pz_{\alpha
j}=\dsum\limits_{i=1}^{n}\dfrac {\widehat{z}_{ij}}
{f_{i\cdot}^{c}f_{\cdot j}^{c} }v_{\alpha i}$. It is clear of
(\ref{eq:eqd71}) and of (\ref{eq:eqd72}) that:

\begin{eqnarray*}
pz_{\alpha j} &=&\dsum\limits_{n=1,v_{\alpha i}>0}^{n}\dfrac{%
\widehat{z}_{ij}}{f_{i\cdot }^{c}f_{\cdot j}^{c}}v_{\alpha
i}+\dsum\limits_{i=1,v_{\alpha
i}<0}^{n}\dfrac{\widehat{z}_{ij}}{f_{i\cdot
}^{c}f_{\cdot j}^{c}}v_{\alpha i} \\
&\leq &\dsum\limits_{n=1,v_{\alpha i}>0}^{n}\frac{\overline{f_{ij}}}{%
f_{i\cdot }^{c}f_{\cdot j}^{c}}v_{\alpha
i}+\dsum\limits_{i=1,v_{\alpha
i}<0}^{n}\frac{\underline{f_{ij}}}{f_{i\cdot }^{c}f_{\cdot
j}^{c}}v_{\alpha i}.
\end{eqnarray*}

Similarly,

\[
pz_{\alpha j}\geq \dsum\limits_{i=1,v_{\alpha i}<0}^{n}\frac{%
\overline{f_{ij}}}{f_{i\cdot }^{c}f_{\cdot j}^{c}}v_{\alpha
i}+\dsum\limits_{i=1,v_{\alpha
i}>0}^{n}\frac{\underline{f_{ij}}}{f_{i\cdot }^{c}f_{\cdot
j}^{c}}v_{\alpha i}
\]

Then $pz_{\alpha j}\in \left[ \underline{\varphi _{\alpha j}},%
\overline{\varphi _{\alpha j}}\right] $, and since
$\underline{\varphi _{\alpha j}}$ and $\overline{\varphi _{\alpha
j}}$ are the projections of some of the tops of the hypercube $Z_{H}^{j}$, then $\underline{%
\varphi _{\alpha j}}$ and $\overline{\varphi _{\alpha j}}$ are
respectively the
minimum and maximum of all the possible projections.\hfill $%
\blacksquare $

\textbf{Theorem 3.} If the hypercube defined by the $i$--th row
profile of interval type is projected on the $\alpha$--th principal
component (in the direction of $u_{\alpha}$), then the maximum and
minimum values are given by the equations (\ref{eq1}) and
(\ref{eq2}) respectively.

\begin{equation}
\underline{\psi _{\alpha i}}=\dsum\limits_{j=1,u_{\alpha j}<0}^{p}\frac{%
\overline{f_{ij}}}{f_{i\cdot }^{c}f_{\cdot j}^{c}}u_{\alpha
j}+\dsum\limits_{j=1,u_{\alpha
j}>0}^{p}\frac{\underline{f_{ij}}}{f_{i\cdot }^{c}f_{\cdot
j}^{c}}u_{\alpha j},  \label{eq1}
\end{equation}

\begin{equation}
\overline{\psi _{\alpha i}}=\dsum\limits_{j=1,u_{\alpha j}<0}^{p}\frac{%
\underline{f_{ij}}}{f_{i\cdot }^{c}f_{\cdot j}^{c}}u_{\alpha
j}+\dsum\limits_{j=1,u_{\alpha
j}>0}^{p}\frac{\overline{f_{ij}}}{f_{i\cdot }^{c}f_{\cdot
j}^{c}}u_{\alpha j}.  \label{eq2}
\end{equation}

\textbf{Proof:} Similar to the previous theorem.\hfill$\blacksquare$

\textbf{Theorem 4.} The classic Correspondence Analysis is a
particular case of the SymCA proposed in Theorem 2 and 3.

\textbf{Proof:} It is evident, because if it is starteed with the
matrix $K=\left( \left[ \underline{k_{ij}},\overline{k_{ij}}\right]
\right)
_{n\times p}$ with $\underline{k_{ij}}=\overline{k_{ij}}$, then $k_{ij}=%
\dfrac{\underline{k_{ij}}+\overline{k_{ij}}}{2}=\underline{k_{ij}}=\overline{%
k_{ij}}$. Then it is gotten for the interval type profiles
$\dfrac{\underline{f_{ij}}}{f_{\cdot j}^{c}}=\dfrac{\overline{f_{ij}}}{%
f_{\cdot j}^{c}}$, and $\dfrac{\underline{f_{ij}}}{f_{i\cdot }^{c}}=\dfrac{\overline{f_{ij}}}{%
f_{i\cdot }^{c}}$, that is to say both are classic profiles.
Therefore algorithm 1 executes a classic Correspondence Analysis,
then the hipercube became a point so the maximum and minimum
coordinates will be the same.\hfill $\blacksquare $

\section{Example}

A contingency table of interval type $K$ of 4 rows and 4 columns in
which appears the eyes color and the hair color of 592 women is
considered , the data table $K$ is given by:

\begin{center}
\begin{tabular}{cc|cccc}
\cline{3-5} &  & \textbf{color} & \textbf{of the} & \textbf{hair} &
\multicolumn{1}{|c}{}
\\ \cline{3-6}
&  & black-h & \multicolumn{1}{|c}{brown-h} &
\multicolumn{1}{|c}{red-h} & \multicolumn{1}{|c|}{blond-h} \\
\hline \multicolumn{1}{|c}{\textbf{color}} &
\multicolumn{1}{|c|}{black-e} & \multicolumn{1}{|c|}{[60,60]} &
\multicolumn{1}{|c|}{[119,123]} & \multicolumn{1}{|c|}{[20,28]} &
\multicolumn{1}{|c|}{[4,7]} \\ \cline{2-6}
\multicolumn{1}{|c}{\textbf{of the}} & \multicolumn{1}{|c|}{brown-e}
& \multicolumn{1}{|c|}{[15,15]} & \multicolumn{1}{|c|}{[50,58]} &
\multicolumn{1}{|c|}{[14,20]} & \multicolumn{1}{|c|}{[5,11]} \\
\cline{2-6} \multicolumn{1}{|c}{\textbf{eyes}} &
\multicolumn{1}{|c|}{green-e} & \multicolumn{1}{|c|}{[5,5]} &
\multicolumn{1}{|c|}{[24,26]} &
\multicolumn{1}{|c|}{[10,12]} & \multicolumn{1}{|c|}{[11,12]} \\
\cline{1-2}\cline{2-6} & \multicolumn{1}{|c|}{blue-e} &
\multicolumn{1}{|c|}{[20,20]} & \multicolumn{1}{|c|}{[70,84]} &
\multicolumn{1}{|c|}{[16,17]} & \multicolumn{1}{|c|}{[90,100]} \\
\cline{2-6}
\end{tabular}
\end{center}

The components of this data table can be interpreted as follows: For
example the entry $[119,123]$ means that the number of brown hair
women with black eyes is between 119 and 123. The entry $[90,100]$
means that the number of blond women with blue eyes is between 90
and 100. The principal plane resultant of the SymCA is presented in
the Figure 1. As it can be seen, the modality ``\textit{black}'' of
the variable hair color is represented by a point, it is coherent
with the information in the data table of this modality, because it
doesn't have any variation (in all the intervals $[a,b]$ of the
first column we have that $a=b$). On the other hand, the modality
``\textit{blond-h}'' and ``\textit{read-h}'' of the variable hair
color are represented by bigger rectangles, that is also coherent
with the information in the data table.

\pict{width=100mm}{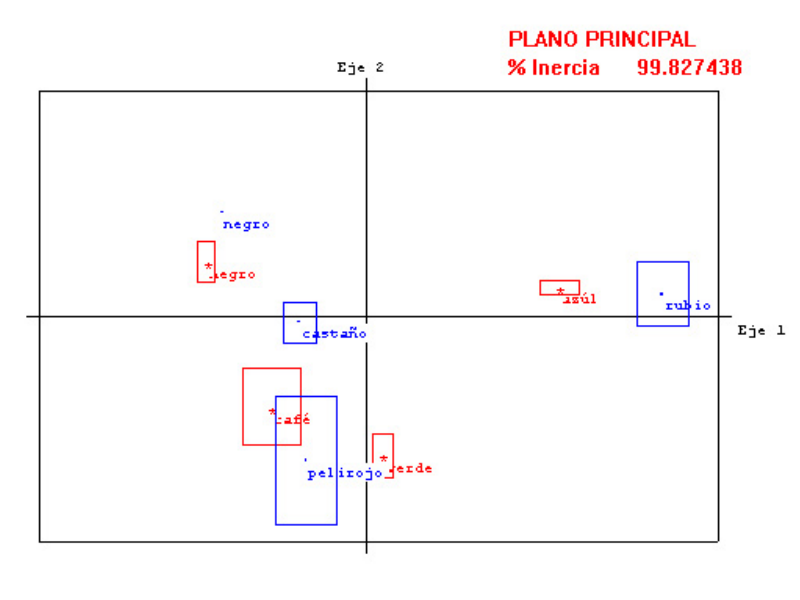}{width=100mm}{graf.pdf}{Graph result of
the SymCA.}{figh1}

\end{document}